\documentclass[prd,amsmath,showpacs,preprintnumbers,superscriptaddress,
               floatfix,twocolumn,letterpaper,
h-physrev4]{revtex4}
\usepackage{times}
\usepackage{graphicx}
\usepackage{amssymb}
\usepackage{mathrsfs}
\newcommand{\ro}{\mathrm}
\newcommand{\ca}{\mathcal}

\begin{document}

\title{Finite-volume matrix Hamiltonian model for a $\Delta \rightarrow N\pi$ system}

\author{J. M. M. Hall} 
\affiliation{Special Research Centre for the Subatomic Structure of
  Matter (CSSM), School of Chemistry and Physics, University of Adelaide 5005,
  Australia} 

\author{A. C.-P. Hsu} 
\affiliation{Special Research Centre for the Subatomic Structure of
  Matter (CSSM), School of Chemistry and Physics, University of Adelaide 5005,
  Australia}

\author{D. B. Leinweber}
\affiliation{Special Research Centre for the Subatomic Structure of
  Matter (CSSM), School of Chemistry and Physics, University of Adelaide 5005,
  Australia}

\author{A. W. Thomas} 
\affiliation{Special Research Centre for the Subatomic Structure of Matter 
  (CSSM), School of Chemistry and Physics, University of Adelaide 5005,
  Australia}
\affiliation{ARC Centre of Excellence for Particle Physics at the Terascale, 
School of Chemistry and Physics, University of Adelaide 5005,
  Australia}

\author{R. D. Young} 
\affiliation{Special Research Centre for the Subatomic Structure of Matter 
  (CSSM), School of Chemistry and Physics, University of Adelaide 5005,
  Australia}
\affiliation{ARC Centre of Excellence for Particle Physics at the Terascale, 
School of Chemistry and Physics, University of Adelaide 5005,
  Australia}

\begin{abstract}

A matrix Hamiltonian model is developed to address 
the finite-volume effects appearing in studies of baryon resonances in 
lattice QCD. 
The Hamiltonian model 
includes interaction terms in a transparent way  
and can be readily generalized to address multichannel problems. 
The eigenvalue equation of the model is exactly solvable and can 
 be matched onto chiral effective field theory. 
The model is investigated in the case of $\Delta\rightarrow N\pi$ scattering. 
A robust method for determining the resonance parameters from lattice QCD 
is developed. It involves constraining the free parameters of the model 
based on the lattice spectrum in question. 
The method is tested in the context of a set of pseudodata, 
and a picture of the model dependence is obtained by examining a variety 
of regularization schemes in the model. 
A comparison is made with the L\"{u}scher method, 
and it is found that the matrix Hamiltonian method is equally robust.  
Both methods are tested in a more realistic scenario, where a background 
interaction corresponding to direct $N\pi\leftrightarrow N\pi$ scattering 
is incorporated into the pseudodata. 
The resulting  extraction of the resonance parameters 
associated with the $\Delta$ baryon 
resonance provides evidence that an effective field theory style 
of approach yields a successful realization of finite-volume effects  
in the context of baryon resonances.

\end{abstract}

\pacs{12.38.Gc 
  12.38.Aw 
  12.39.Fe 
}
\maketitle
\preprint{ADP-13-10/T830}

\section{Introduction}
\vspace{-1mm}

The behaviour of the excited states of baryons 
represents an important research topic 
in nuclear and particle physics. The extraction of 
resonance parameters (e.g. masses and widths) provides insight into the 
scattering behaviour of hadronic interactions. 
Lattice QCD constitutes a principal tool for the analysis of nonperturbative 
physics; however, observables must be computed in a finite volume. 
Finite-volume models of hadronic interactions represent a 
valuable avenue for investigating the   
intrinsic features of lattice QCD. They assist in the interpretation 
of discrete eigenstates in relation to the asymptotic behaviour of the 
$S$ matrix measured in experiments. 
Recent advances in lattice QCD have enabled the
extraction of the lowest-lying excitation energies 
 \cite{Edwards:2011jj,Engel:2010my,Mahbub:2010rm,Menadue:2011pd,Mohler:2012nh}. However,  
the extraction of resonance parameters 
 from multihadron interactions presents  
an ongoing challenge for current research 
\cite{He:2005ey,Bernard:2008ax,Bernard:2009mw,Doring:2011nd,Doring:2011vk,Roca:2012rx,Giudice:2012tg,Kreuzer:2012sr,Albaladejo:2012jr,Dudek:2012xn,Gockeler:2012yj}. 

In this article, a finite-volume matrix Hamiltonian model is introduced, which 
is exactly solvable. A matrix is constructed, 
and its entries are populated with energies corresponding to different 
momentum states of the interaction channel(s). 
The solution of the eigenvalue equation takes the form of a  
one-loop renormalization formula for a bare resonance energy, reminiscent 
of finite-volume chiral effective field theory ($\chi$EFT). 
The Hamiltonian can then be matched onto the corresponding $\chi$EFT formula 
by choosing the form of the interaction in the model. 
A property of the formalism is the ability to reproduce the 
 ``avoided level crossing'' observed in lattice QCD 
calculations, while also being able to reproduce
 the appropriate continuum limit of $\chi$EFT.  
Thus the physical behaviour of the energy 
eigenvalues near a resonance due to the mixing of the particle states can 
be estimated. 
Furthermore, the Hamiltonian method lends itself to an intuitive generalization 
in addressing multichannel problems, with the inclusion of additional 
degrees of freedom, corresponding to the new interactions. 

The matrix Hamiltonian formalism is applied to $\Delta \rightarrow N\pi$ decay, 
and its energy spectrum is generated for a range of finite box sizes, $L$. 
A method for extracting the resonance parameters from a lattice QCD 
spectrum is then postulated, which involves fitting the free 
parameters of the model. The method is tested by generating 
a set of pseudodata from the model itself. 
The pseudodata are analysed with an alternative version 
of the model and a selection of different regularization parameters. 
This provides an indication of the model dependence in correctly obtaining 
 the resonance position. 
The success of the 
identification of the phase shift is compared to that of the L\"{u}scher method 
\cite{Luscher:1990ux}. 
This serves to establish a benchmark for the Hamiltonian model in obtaining 
the resonance position. 
It is found that the matrix Hamiltonian model, as applied to 
$\Delta\rightarrow N\pi$ decay, is of comparable accuracy to the 
L\"{u}scher method.

\vspace{-2mm}

\section{Scattering theory from effective field theory}
\vspace{-1mm}

In continuum scattering theory, below inelastic thresholds, 
amplitudes of the partial-wave decomposition of the wavefunction, 
which are complex valued, may be converted into 
a real-valued number called the phase shift, $\delta_l(k)$, 
by exploiting the conservation of angular momentum.
The phase shift, which depends on the orbital angular momentum $l$ 
and external momentum $k$,  
is related to the total cross section of a scattering process, via
\begin{align}
\sigma(k) &= \int\!\ro{d}\Omega |f(k,\theta)|^2,\\
\mbox{where}\quad f(k,\theta) &= 
\sum_l(2l+1) P_l(\cos\theta)\frac{e^{i\delta_l(k)}\sin\delta_l(k)}{k},
\end{align}
where $P_l$ is the $l$th Legendre polynomial. 
The external momentum of a particle with mass $m$ is related 
to the external energy 
by: $k = \sqrt{E^2-m^2}$. 
In the case of $P$-wave scattering ($l=1$), 
the total cross section simplifies to the form
\begin{equation}
\sigma(k) = 4\pi \frac{3\sin^2\delta(k)}{k^2},
\end{equation}
where $\delta(k) \equiv \delta_1(k)$. 
The energy at which the phase shift passes 
through $90^\circ$ is the resonance energy, $E_{\ro{res}}$. 

\begin{figure}[tp]
\centering
\includegraphics[height=80pt]{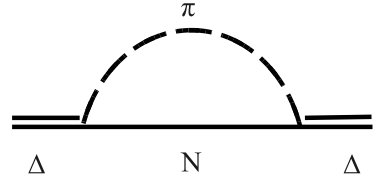}
\caption{The leading-order one-pion loop contribution 
to the resonance energy $E$ associated with $\Delta\rightarrow N\pi$ decay. 
All charge conserving transitions are implicit.}
\label{fig:loop}
\end{figure}

The investigation of $\Delta\rightarrow N\pi$ decay is of particular interest.  
It provides insight into the $\Delta$ resonance of $N\pi$ scattering; 
an archetypal example of a baryon resonance.  
The on-shell $t$ matrix  
associated with $N\pi$ scattering from a $\Delta$ baryon, 
$T \equiv t(k,k;E^+)$,  
can be obtained by solving the Lippmann-Schwinger equation, 
and the phase shift is directly related to the $t$ matrix, via 
\begin{align}
\label{eqn:tprop}
T &= \frac{g_{\Delta N}^2(k)}{E - \Delta_0 - \Sigma_{\Delta N}(k)}\\
&= -\frac{1}{\pi k E} e^{i\delta(k)}\sin\delta(k). 
\label{eqn:tmat}
\end{align}
The quantity $\Delta_0$ represents the bare value of the resonance energy, 
which becomes renormalized by the one-pion loop integral, $\Sigma_{\Delta N}(k)$ 
(shown in 
Fig.~\ref{fig:loop}), and 
may be chosen such that $E_{\ro{res}}$ matches the phenomenological value, $E_{\ro{res}}^{\ro{phys}} = 292$ MeV. 
By working in the heavy-baryon approximation, 
the kinetic energy of the nucleon is 
neglected in this simple model, 
and calculations may be performed relative to the nucleon mass. 
Thus, the external energy may be expressed in terms of the pion energy, 
$E=\omega_\pi(k)$.

The coupling $g_{\Delta N}(k)$ takes the following form, as obtained from 
$\chi$EFT,  
\begin{equation}
\label{eqn:coup}
g_{\Delta N}^2(k) = \chi_\Delta \frac{2}{\pi}\frac{k^2\,u^2(k)}{\omega_\pi(k)}, 
\end{equation}
where the coefficient $\chi_\Delta$  
is defined in terms of known phenomenological 
parameters: $f_\pi=92.4$ MeV and 
$\mathcal{C}=-1.52$, as derived from the SU$(6)$ flavour-symmetry relation 
\begin{equation}
\chi_\Delta = \frac{3}{32\pi f_\pi^2} \frac{2}{9}\ca{C}^2.
\end{equation}
The parameters $f_\pi$ and $\ca{C}$ occur in the first-order interaction 
Lagrangian of chiral perturbation theory 
($\chi$PT) describing the $\Delta \rightarrow N\pi$ interaction 
\cite{Jenkins:1991ne,Jenkins:1990jv,Jenkins:1991ts,Labrenz:1996jy,WalkerLoud:2004hf,Wang:2007iw}
\begin{equation}
\ca{L}_{\chi\ro{PT}} = -\frac{\ca{C}}{2 f_\pi} \bar{N} T^a \gamma^{\mu\nu}\Delta_\nu \partial_\mu \pi^a + \ro{h.c.},
\end{equation}
where $T^a$ are the relevant $2\times 4$ isospin transition matrices 
\cite{Pascalutsa:2005vq}.

Note that a finite-range regulator $u(k)$ has been introduced 
into the coupling 
term of Eq.~(\ref{eqn:coup}). This serves to control the otherwise ultraviolet 
divergent loop integral and, in addition, will be used in Sec.~\ref{sec:fin}
 to establish 
a finite interaction range in the Hamiltonian model. 
For more detailed discussions of finite-range regularization, see 
Refs.~\cite{Young:2002ib,Leinweber:2003dg,Leinweber:2005xz,Hall:2010ai,Hall:2011en}.

\begin{figure}
\begin{center}
\includegraphics[height=0.950\hsize,angle=90]{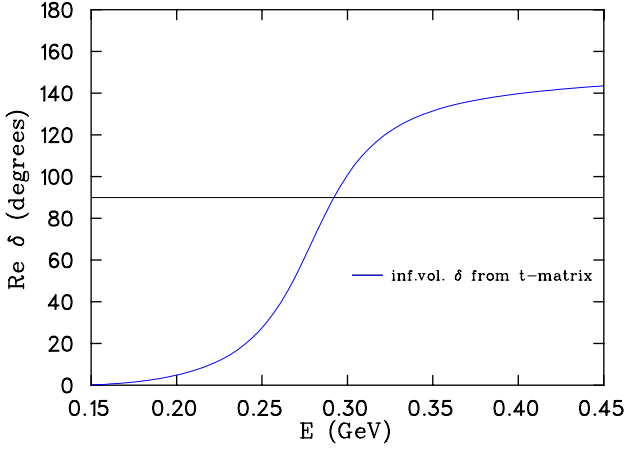}
\vspace{-11pt}
\caption{(color online). The infinite-volume phase shift $\delta$ associated with elastic $N\pi$ scattering via a $\Delta$ baryon intermediate state, plotted against the external pion energy $E$ (where $E_\ro{tot} = M_N + E$), as calculated from the on-shell $t$ matrix of Eq.~(\ref{eqn:tmat}).}
\label{fig:dvsEinf}
\end{center}
\end{figure}

The leading-order pion loop integral 
shown in Fig.~\ref{fig:loop} takes the form
\begin{align}
\Sigma_{\Delta N}(k) &= \int_0^\infty\!\!\ro{d} 
k'\frac{k'\,^2 g_{\Delta N}^2(k'\, )}
{\omega_\pi({k} ) - \omega_\pi(k'\, )-i\epsilon} \\
&= \chi_\Delta \frac{2}{\pi}\int_0^\infty\!\!\ro{d} 
k'\frac{k'\,^4\, u^2(k'\, )}
{\omega_\pi(k'\, )\,[\omega_\pi({k} ) - \omega_\pi(k'\, ) -i\epsilon]},
\label{eqn:int}
\end{align}
where $\mathcal{P}$ indicates that a principal value integral must  
be performed. 
By using Eq.~(\ref{eqn:tmat}), it is straightforward to solve for $\delta(k)$. 
A plot of the phase shift against external energy, $E = \omega_\pi(k)$, 
is shown in 
Fig.~\ref{fig:dvsEinf}, which takes the form of a Breit-Wigner-like curve.

\section{Finite-volume Hamiltonian model}
\label{sec:fin}
\vspace{-1mm}

In the following finite-volume heavy-baryon model, a  
matrix Hamiltonian is defined to include the interaction  
between the $\Delta$ baryon and the pion-nucleon system. 
The rows and columns of $H$ represent the total three-momentum 
values available to the pion 
($k_1 = 2\pi/L,\,k_2 = 4\pi/L\ldots$). 
In the simplest version of the model, 
the $N\pi$ states only couple to the $\Delta$ baryon.  

The Hamiltonian may be written as the sum of free and interaction 
Hamiltonians, $H = H_0 + H_I$. 
The free Hamiltonian, $H_0$, takes the form
\begin{equation}
\label{eqn:H0}
 H_0  = 
\begin{pmatrix}
\Delta_0& 0 & 0 &\cdots\\
0 & \omega_\pi(k_1)
& 0 &\cdots\\
0 & 0 &\omega_\pi(k_2)
\\
\vdots &\vdots & & \ddots     \end{pmatrix},\\
\end{equation}
where $\omega_\pi(k_n)\equiv\sqrt{k_n^2 + m_\pi^2}$ is the pion energy, 
and the pion momenta at finite volume take only the discrete 
values available in a box of length $L$, obeying the condition:
$k_n^2 = \left(\frac{2\pi}{L}\right)^2 n$, for a squared integer 
$n\equiv n_x^2 + n_y^2 + n_z^2$.
$H_0$ has nonzero diagonal entries representing 
the noninteracting energy of a pion at certain values of momentum.    
The nucleon recoil energy vanishes in the heavy-baryon limit.  
The first element of $H_0$ 
represents the bare resonance mass relative to the nucleon mass, 
$\Delta_0$. The value of $\Delta_0$ is chosen so that the resonance 
position takes the phenomenological value $292$ MeV, 
in the infinite-volume case in Eq.~(\ref{eqn:tprop}).

 The interacting Hamiltonian $H_I$ contains couplings $g_{\Delta N}$  
in the top row and left-most column,
\begin{equation}
\label{HI}
 H_I  = \begin{pmatrix}
0& g^{\ro{fin}}_{\Delta N}(k_1) &  g^{\ro{fin}}_{\Delta N}(k_2) &\cdots\\
  g^{\ro{fin}}_{\Delta N}(k_1)  & 0 &0 &\cdots\\
 g^{\ro{fin}}_{\Delta N}(k_2)  & 0 &0 &\cdots\\
\vdots & \vdots& \vdots&\ddots     \end{pmatrix}.
\end{equation}
This represents the interaction between the bare $\Delta$ baryon and the 
$N\pi$ system over a range of momentum values. 
The couplings are chosen to 
take a similar form to that of Eq.~(\ref{eqn:coup})  
but contain additional factors relevant to the normalization of the 
discrete momenta in a finite-volume box, 
\begin{equation}
\label{eqn:gfin}
g^{\ro{fin}}_{\Delta N}(k_n) = 
\sqrt{\frac{C_3(n)}{4\pi}}\left(\frac{2\pi}{L}\right)^{3/2}
g_{\Delta N}(k_n). 
\end{equation}
Note that the  
cubic symmetry of the system means that 
each nonzero $n$ need occur only once in the rows and columns of $H$, 
so long as the finite-volume couplings $g^{\ro{fin}}_{\Delta N}(k_n)$ 
contain the appropriate normalization factor $\sqrt{C_3(n)}$, 
where $C_3(n)$ represents 
the number of ways of summing the squares of three integers to equal $n$. 

For the purposes of this model, $u(k_n)$ is chosen to take 
the form of a dipole regulator,  
with a regularization scale of $\Lambda = 0.8$ GeV, 
\begin{equation}
\label{eqn:ureg}
u(k_n) = {\left(1 + {\frac{k_n^{2}}{\Lambda^{2}}} \right)}^{-2}.
\end{equation}
However, in Sec.~\ref{sec:luescher}, 
the associated model dependence of the choice of regulator on 
the extraction of the resonance parameters from the model is investigated  
and a variety of regularization schemes are examined. 

\begin{figure}[tp]
\begin{center}
\includegraphics[height=0.95\hsize,angle=90]{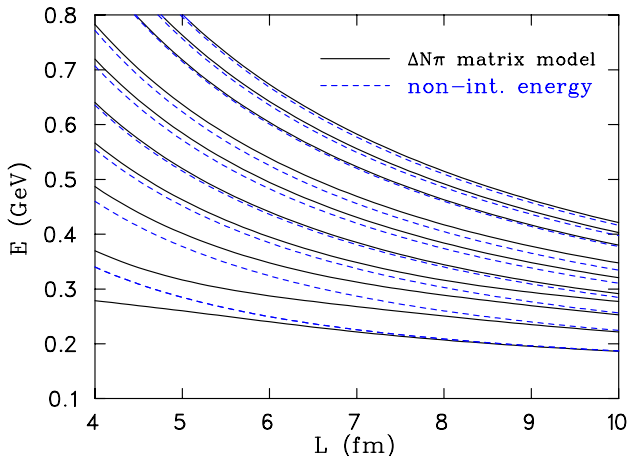}
\vspace{-11pt}
\caption{(color online). The lowest-lying energy levels from the $\Delta N \pi$ model, described in Eqs.~(\ref{eqn:H0})$-$(\ref{eqn:ureg}) 
(solid lines), and the corresponding noninteracting energies (dotted lines).}
\label{fig:EvsLfina}
\end{center}
\end{figure}

The eigenvalues of the Hamiltonian matrix 
correspond to the different energy levels 
in a finite volume, and may be calculated for a range of $L$ values. 
The ten lowest energy levels as a function of $L$ 
are shown in Fig.~\ref{fig:EvsLfina}, with 
the noninteracting energies shown as dotted lines.

The Hamiltonian model, by construction, matches 
onto the $\chi$EFT formula for the renormalization of the 
resonance mass in a finite volume. 
The eigenvalue equation of the Hamiltonian, 
$\ro{det}(H - \lambda \mathbb{I}) = 0$, can be solved exactly, 
and takes the form
\begin{align}
\lambda &= \Delta_0 - \sum_{n=1}^\infty 
\frac{(g_{\Delta N}^{\ro{fin}}(k_n))^2}{\omega_\pi(k_n)-\lambda} \\
&= \Delta_0 - \frac{\chi_\Delta}{2\pi^2}\left(\frac{2\pi}{L}\right)^3 
\sum_{n=1}^\infty C_3(n)
\frac{k_n^2 u^2(k_n)}{\omega_\pi(k_n)[\omega_\pi(k_n)-\lambda]}.
\label{eqn:sum}
\end{align}
A comparison with Eq.~(\ref{eqn:int}) shows that the sum term takes the same 
form as the loop integral, upon the restoration of the continuum integral 
\begin{equation}
\left(\frac{2\pi}{L}\right)^3 \sum_{n=1}^\infty C_3(n)\rightarrow \int\!\ro{d}^3 k.
\end{equation} 
This correspondence is a consequence of
the judicious choice of the form of the coupling in the finite-volume 
Hamiltonian $g^{\ro{fin}}_{\Delta N}$.

The subtle, yet important difference between Eqs.~(\ref{eqn:int}) 
and (\ref{eqn:sum}) is that the energy eigenvalue $\lambda$ 
appears on both sides of the equation derived from the Hamiltonian. 
The presence of $\lambda$ in the denominator, whose solution must be 
finite, results in the avoided level crossing observed in 
lattice QCD, where the resonance energy corresponds to maximal 
mixing of the single-particle and multiparticle states. 
That is, the correct quantum mechanical behaviour 
of the multiparticle system is directly built into the Hamiltonian 
model. This is the finite-volume analogue of the principal value integral 
in the continuum theory. 

The Hamiltonian model may be extended to include a 
direct $N\pi\leftrightarrow N\pi$ interaction, 
which could be viewed as an approximation to the Chew-Low interaction. 
For example, one can add additional terms in the interaction matrix, $H_I$:
\begin{equation}
\label{eqn:CL}
g_{CL}(k_i,k_j) = -g_{N\pi}(k_i)g_{N\pi}(k_j)
\frac{1}{\sqrt{\omega_\pi(k_i)\omega_\pi(k_j)}}.
\end{equation}
In the cloudy bag model \cite{Theberge:1980ye,Thomas:1981vc}, the couplings 
$g_{N\pi}$ and $g^{\ro{fin}}_{\Delta N}$ are related by the equation  
$g_{N\pi}(k_i) = \kappa g_{\Delta N}^{\ro{fin}}(k_i)$ with  
a ratio $\kappa = \sqrt{25/18}$ 
estimated from SU$(6)$ 
symmetry. 
In the case of the $t$ matrix, the Lippmann-Schwinger 
equation with a rank-$2$ separable potential can be solved as per Mongan 
\cite{Mongan:1969dc}.

This additional interaction  
will be used to test the model independence in the procedure of extracting 
the phase shift in Sec.~\ref{sec:luescher}. 
The comparison of two models, one with extra background 
interactions of the form of Eq.~(\ref{eqn:CL}) and one without the extra 
interactions, 
will provide insight 
into the robustness of the identification of resonance parameters from 
an effective model.

\section{Phase shift extraction from the lattice}
\vspace{-1mm}

The identification of resonance parameters in a finite volume 
is important for interpreting lattice QCD results in the light of 
experiment. 
The connection between the energy spectra of lattice QCD simulations  
and the asymptotic states of the $S$ matrix 
represents a current challenge for research in finite-volume 
hadron spectroscopy 
\cite{He:2005ey,Liu:2005kr,Bernard:2012bi,Doring:2012eu,Hansen:2012tf}. 

The features of the 
Hamiltonian model are ideally suited to the extraction of 
resonance parameters from lattice QCD energy spectra. 
Not only is the construction of the Hamiltonian transparent 
and intuitive in terms of the potentials one chooses to include, 
but the formalism may be easily generalized to include multiple channels. 
By generating energy eigenvalues from a Hamiltonian model, 
such as those shown in Fig.~\ref{fig:EvsLfina} for the $\Delta \rightarrow N\pi$ 
interaction, the free parameters of the model may be tuned 
to fit the energy spectrum of a lattice QCD calculation. 
In the case of the $\Delta \rightarrow N\pi$ model, this entails 
minimizing the chi-square for the parameters $\chi_\Delta$ and $\Delta_0$. 
The resonance energy and phase shift may then be extracted by using the fitted 
 values of the parameters, input into Eq.~(\ref{eqn:tmat}). 
An infinite-volume phase shift, much like that of Fig.~\ref{fig:dvsEinf},  
is thus 
obtained, except that the underlying parameters have been extracted  
directly from lattice QCD results.

\section{Comparison with the L\"{u}scher method}
\label{sec:luescher}
\vspace{-1mm}

In order to demonstrate the success of the Hamiltonian method 
in reproducing accurate phase shifts from the finite-volume 
energy spectrum of the model, a comparison is made with the 
well-known L\"{u}scher method \cite{Luscher:1990ux}. 
L\"{u}scher's formula describes a fixed relationship between the 
scattering phase shift $\delta$ and the momentum corresponding 
to the $j$th energy level in a 
finite volume
\begin{equation}
\label{lf}
\delta(k_j;L) = j\,\pi - \phi\left(\frac{k_j L}{2\pi}\right).
\end{equation}
The kinematic 
function $\phi(q)$ takes the form of a three-dimensional Zeta-like 
function (which must be regularized in an appropriate fashion), 
defined in terms of dimensionless lattice momenta $q \equiv k L/(2\pi)$, 
\begin{align}
\label{phi}
\phi(q)&=-\ro{arctan}\left(\frac{\pi^{3/2} q}{\mathcal{Z}_{00}(1,q^2)}\right),
\\
\mathcal{Z}_{00}(1,q^2)&= \frac{1}{\sqrt{4\pi}}\sum_{\vec{n}\in\mathbb{Z}^3}
\frac{1}{\vec{n}^2-q^2}.
\end{align}
The momenta corresponding to a lattice energy spectrum   
may be input into L\"{u}scher's formula  
to obtain volume-dependent ``phase shifts'' $\delta(k_j;L)$. 

\begin{figure}[tp]
\includegraphics[height=0.950\hsize,angle=90]{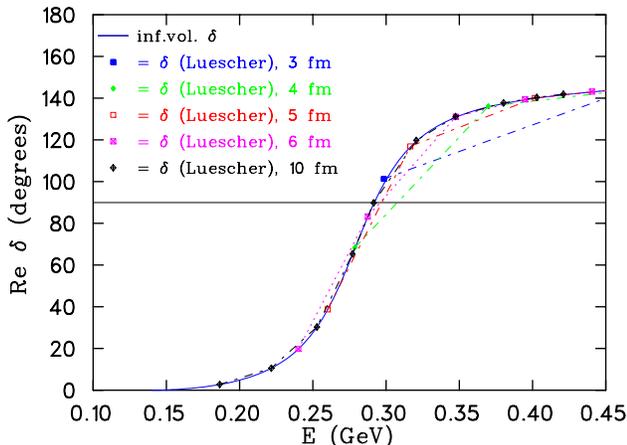}
\vspace{-11pt}
\caption{\footnotesize{(color online). Estimates of the phase shift obtained from L\"{u}scher's formula applied to the $\Delta \rightarrow N\pi$ model for a range of $L$ values. The infinite-volume phase shift is also shown as a solid line. }}
\label{fig:dvsEL}
\end{figure}

As an example, consider the application of the L\"{u}scher method 
to the energy spectrum of the $\Delta \rightarrow N\pi$ model, treating the spectrum 
as pseudodata. By 
taking the momenta associated with the energy eigenvalues, shown in 
Fig.~\ref{fig:EvsLfina}, as inputs, and choosing a particular box size $L$, 
the finite-volume estimates of the phase shifts may be compared 
with the known infinite-volume phase shifts, 
as shown in Fig.~\ref{fig:dvsEinf}. 
The result is plotted on the same axes, in Fig.~\ref{fig:dvsEL},  
for a range of box sizes. 
Both the infinite-volume phase shift and the Hamiltonian energy spectrum 
are generated using a dipole regulator with $\Lambda = 0.8$ GeV. 
Figure~\ref{fig:dvsEL} shows that the finite-volume corrections 
associated with applying the L\"{u}scher method to 
a finite-volume effective Hamiltonian model are insignificant 
for $L\geq 3$ fm at the physical pion mass. 
It will be demonstrated that the matrix 
Hamiltonian method of obtaining a phase shift from finite-volume 
energy levels is comparable with the L\"{u}scher method.

Note that, in using L\"{u}scher's method, 
an interpolation function must be chosen 
in order to obtain the pole position from the phase shift. 
The Breit-Wigner 
curve represents a suitable functional form,
\begin{equation}
e^{i\delta(k)}\sin\delta(k) = \frac{\tilde{\Gamma}(k)/2}{E-E_{\ro{res}}-i\tilde{\Gamma}(k)/2},
\end{equation}
where the width contains a cubic momentum factor, which 
yields the correct phase-space dependence for $l=1$  
\cite{Blatt},
\begin{equation}
\tilde{\Gamma}(k) = \frac{k^3}{k_{\ro{res}}^3}\Gamma. 
\end{equation}

To illustrate the comparison between L\"{u}scher's method and the
 matrix Hamiltonian method, the pseudodata obtained from the 
$\Delta \rightarrow N\pi$ model will again be used. 
A modified version of the $\Delta \rightarrow N\pi$ model is constructed 
by replacing the regulator function $u(k_n)$ with a different 
form of regularization. Firstly, a Gaussian regulator with 
a variety of values for the regularization scale $\Lambda$ will be 
considered. 
Using the $\Delta \rightarrow N\pi$ model, the closest two energy eigenvalues
 to the value of $E_{\ro{res}}$, as estimated by L\"{u}scher's formula, 
are chosen in order to 
constrain $\chi_\Delta$ and $\Delta_0$. 

\begin{figure}[tp]
\begin{center}
\includegraphics[height=1.0\hsize,angle=90]{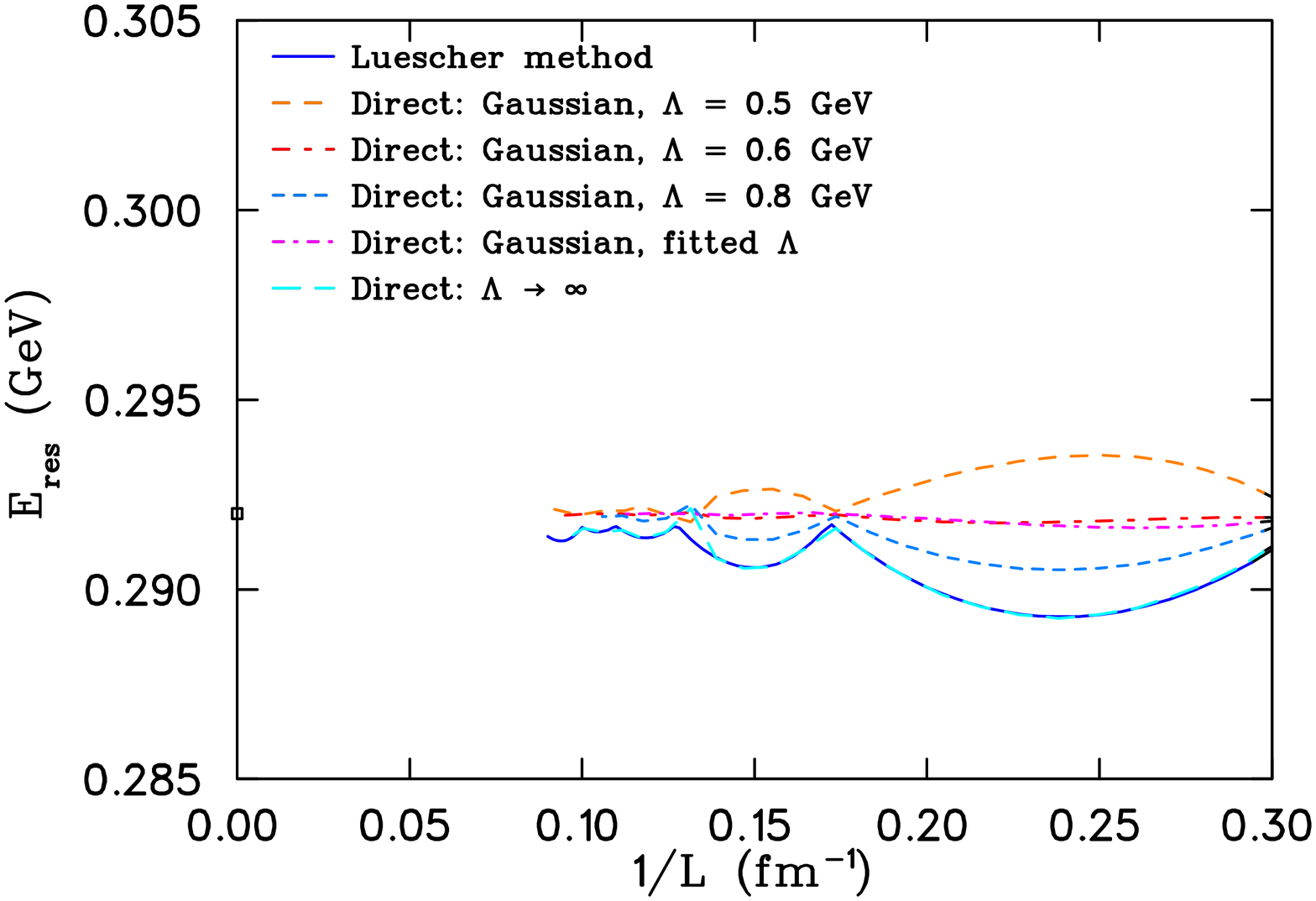}
\vspace{-11pt}
\caption{(color online). The resonance energy $E_{\ro{res}}$ plotted against $1/L$. 
The experimental value is marked with a square. The results from fitting pseudodata with a different regulator, a Gaussian with $\Lambda = 0.5,\,0.6$ and $0.8$ GeV, or treating $\Lambda$ as a fit parameter, are plotted. The result of effectively removing the regulator (i.e. $\Lambda\rightarrow\infty$) is also shown. For comparison, the approach using L\"{u}scher's method is marked with a solid line.} 
\label{fig:EresvsLinv}
\end{center}
\vspace{-1mm}
\begin{center}
\includegraphics[height=1.0\hsize,angle=90]{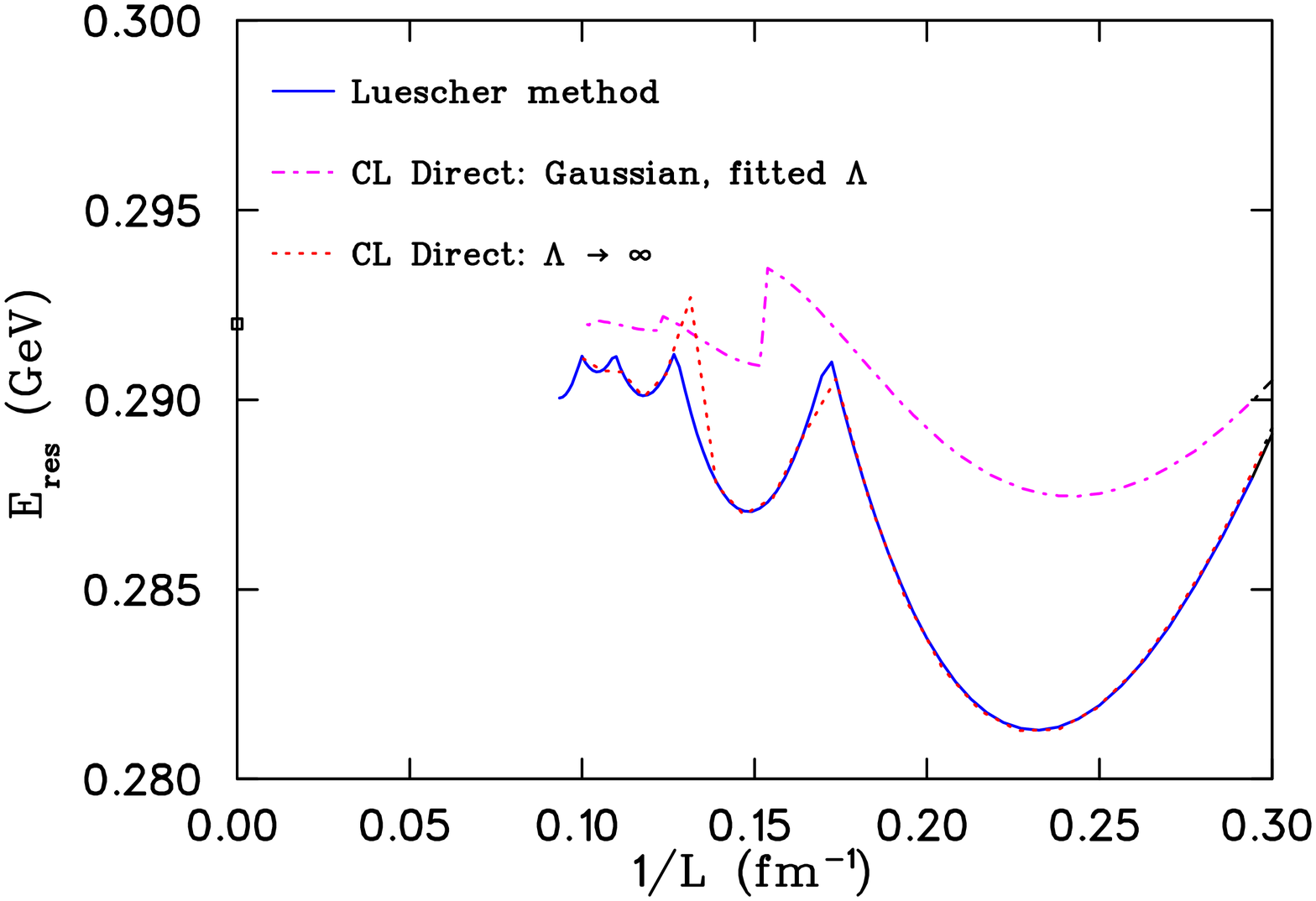}
\vspace{-11pt}
\caption{(color online). The resonance energy, $E_{\ro{res}}$, plotted against $1/L$, resulting from fitting pseudodata containing the nonresonant background interaction of Eq.~(\ref{eqn:CL}). 
The experimental value is marked with a square. The result from fitting pseudodata with a Gaussian regulator is plotted, treating $\Lambda$ as a fit parameter. The result corresponding to  $\Lambda\rightarrow\infty$ is also shown. L\"{u}scher's approach is marked with a solid line.} 
\label{fig:EresvsLinvCL}
\end{center}
\end{figure}

The  behaviour of the resonance energy $E_{\ro{res}}$ may be   
plotted as a function of $1/L$, as shown in Fig.~\ref{fig:EresvsLinv}. 
  A key observation is that 
the periodic deviations in the extraction of $E_{\ro{res}}$ arise from the 
fact that, at certain volumes, 
an energy level may or may not lie near the resonance position 
(where $\delta = 90^\circ$). That is, they are simply artefacts 
associated with the choice of interpolation. 
For box sizes where a discrete 
energy eigenvalue does lie near $E_{\ro{res}}$, the experimental value 
is attained within a few MeV. This can provide a guide in choosing 
suitable box sizes for reliably reproducing the position 
of the $\Delta$ resonance.

It is also worth noting that the use of a Gaussian 
regulator, with $\Lambda \simeq 0.6$ GeV, 
provides a much more reliable reconstruction of the 
resonance position from the pseudodata than a  
 low or high value of $\Lambda$. 
This is a consequence of a close match between the 
underlying regulator of the pseudodata, and the Gaussian regulator at this 
value of $\Lambda$. It provides evidence that a choice of regulator 
that correctly reflects the underlying phenomenology leads to a more 
stable identification of the resonance position. 
Nevertheless, even in the case of $\Lambda\rightarrow\infty$, 
which essentially corresponds to a dimensional regularization-like  
result \cite{Leinweber:2003dg}, the matrix Hamiltonian method 
is just as successful as the L\"{u}scher method.

Note that in Fig.~\ref{fig:dvsEL}  use of the L\"{u}scher method 
results in an excellent agreement with the infinite-volume 
phase shift. However, at small 
volumes, the sparse distribution of energy eigenvalues leads to a difficulty 
in extracting the behaviour at $\delta = 90^\circ$. This accounts for 
the model dependence apparent in Fig.~\ref{fig:EresvsLinv}. 

One might argue that the analysis of the pseudodata using a 
similar model that includes only the $\Delta \rightarrow N\pi$ 
couplings is ideally suited to the reconstruction method proposed above. 
In a more realistic scenario, a new pseudodata set is  
generated from a Hamiltonian model that now includes   
the effective contact interaction, introduced in Eq.~(\ref{eqn:CL}).  
Again, a dipole regulator with $\Lambda = 0.8$ GeV is used. 
The energy levels of the new pseudodata are fitted to the more 
basic model  
using a Gaussian regulator, and the regularization scale, $\Lambda$,  
is treated as an additional fit parameter. 
The  behaviour of the resonance energy versus $1/L$ 
is displayed in Fig.~\ref{fig:EresvsLinvCL}. 
The fitted values of $\Lambda$ in Fig.~\ref{fig:EresvsLinv} are 
recalculated at each value of $L$, but they are typically close to 
$0.6$ GeV, whereas in Fig.~\ref{fig:EresvsLinvCL} the values are 
closer to $0.5$ GeV. The tendency for slightly smaller fitted values of 
$\Lambda$ in Fig.~\ref{fig:EresvsLinvCL} simply reflects the 
compensation of the regulator for the missing background interaction. 
For all regularization methods tested, the presence of a 
background interaction also causes an exaggeration in the 
periodic deviations of $E_{\ro{res}}$ from interpolation. 
In all cases, the deviations are much larger due to the effect of the 
nonresonant background interaction. Nevertheless, a fairly 
reliable extraction of $E_{\ro{res}}$ is possible, if the energy eigenvalues 
lie close enough to the resonance energy. 
Most notably, there is significant improvement in the identification 
of $E_{\ro{res}}$ using the matrix Hamiltonian method if 
$\Lambda$ is treated as a fit parameter. 
This is remarkable, as it shows that the bare resonance mass, the 
coefficient of the interaction and the finite-range scale $\Lambda$ 
are collectively able to compensate for the missing $N\pi\leftrightarrow N\pi$ 
interactions; the 
underlying interactions are simply incorporated into the 
fit degrees of freedom. 

\section{Analysis of the eigenvectors}
\vspace{-1mm}

An analysis of the eigenvectors  
serves to emphasise a key feature of the Hamiltonian model. 
While only a single local operator describing the $\Delta$ resonance is 
included in the model, 
the interactions drive 
a significant coupling to all nearby lattice eigenstates. 
 That is, one observes the generation of quantum 
mechanical admixtures of states from the interactions. 

\begin{figure}[tp]
\includegraphics[height=0.930\hsize,angle=90]{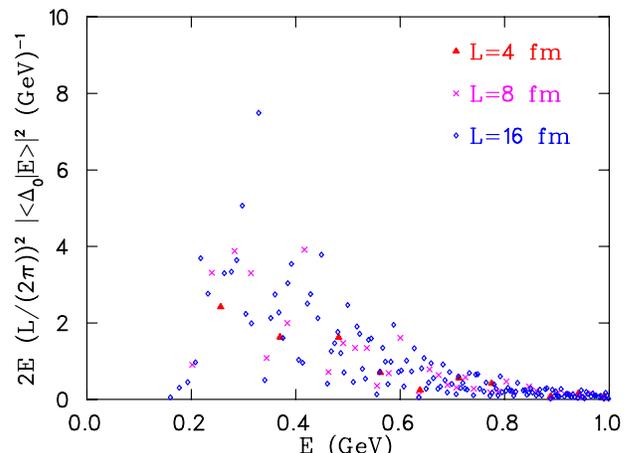}
\vspace{-11pt}
\caption{\footnotesize{(color online).  The contribution of the eigenvectors 
to the bare $\Delta$ state in a finite volume, for a range of box sizes. 
The finite-volume Hamiltonian used includes the $N\pi\leftrightarrow N\pi$ 
interaction terms. }}
\label{fig:evec}
\end{figure}

Consider the square of the 
overlap of each eigenvector with the bare state $|\Delta_0\rangle$, with a 
volume-dependent coefficient calculated from the following Jacobian:
\begin{align}
\ro{d}k^2 &= \left(\frac{2\pi}{L}\right)^2\ro{d}n \\
&= \ro{d}E^2 = 2 E\, \ro{d}E,\\
&\Rightarrow \int\!\ro{d}n|\langle\Delta_0|E_n\rangle|^2 
= \int\!\ro{d}E\,2E\,\left(\frac{L}{2\pi}\right)^2|\langle\Delta_0|E\rangle|^2. 
\end{align}
The result may 
be plotted as a function of external energy $E$.  
The peak of the resulting 
curve should lie at the value of the renormalized mass of the $\Delta$ baryon 
relative to the nucleon,
 $292$ MeV. 
The result for the matrix Hamiltonian model, using a  
dipole regulator with $\Lambda = 0.8$ GeV, 
and including the Chew-Low-like background 
interactions from Eq.~(\ref{eqn:CL}), 
is shown in 
Fig.~\ref{fig:evec}. 
The magnitude of the contribution 
from the eigenvector spectrum to $\Delta_0$, $|\langle\Delta_0|v\rangle|^2$, 
is shown for box sizes of $4$, $8$ and $16$ fm. 
Over a broad range of energies, which dominate across the width of the 
resonance, it is evident that no single eigenstate, by itself, 
can be identified as either a local or a multihadron state. 
That is, the local, noninteracting $\Delta_0$ state couples to a range of 
low-lying states. This illustrates how early ideas from the heavier 
quark-mass domain, in distinguishing 
single-particle and multiparticle states, 
are no longer applicable in the light-quark 
regime of modern lattice simulations.

\vspace{-2mm}
\section{Conclusion}
\vspace{-1mm}

Finite-volume models provide key insights into the intrinsic 
attributes of lattice QCD. By examining the behaviour of 
resonances and multiparticle states, one is able to identify more 
precisely the impact on a calculation associated with finite-volume effects. 

By introducing a matrix Hamiltonian method, the finite-volume behaviour 
near a resonance is reproducible. 
The avoided level crossings that indicate 
nontrivial mixing between single-particle and multiparticle 
states are also present 
in the model. 
Because the Hamiltonian is exactly solvable, the eigenvalue equation 
may be matched onto chiral effective field theory simply by a choice 
of coupling parameter for the potential. 
The Hamiltonian model can also be easily generalized to include multiple 
channels, in order to address more difficult scattering problems. 

As a simple, introductory example, 
the $\Delta$ baryon resonance of $N\pi$ scattering 
was considered, and a method for extracting the phase shift 
from lattice QCD was developed. By matching the energy eigenstates of the model 
to those of a lattice calculation, the free parameters of the model 
were constrained, and their values were input into a scattering $t$ matrix 
equation, which uses chiral effective field theory. 
  Initially, 
only the leading-order one-pion loop integral contribution to the resonance 
energy was included. 

The method was tested in the context of a set of pseudodata generated 
by the $\Delta \rightarrow N\pi$ model. 
By considering a variety of different regularization 
scales, an indication of the model dependence in the calculation was 
found. A comparison was made with the well-known L\"{u}scher method, 
and it was discovered that the matrix Hamiltonian method was at least 
comparable in reliability, with improvement when the regulator 
is a good approximation of the underlying phenomenology.

The extraction of the resonance 
parameters was investigated in a more realistic context, where 
nonresonant background interaction terms (corresponding to direct 
$N\pi\leftrightarrow N\pi$ scattering) were incorporated into the pseudodata. 
It was found that the $\Delta \rightarrow N\pi$ model was able to 
improve upon the L\"{u}scher method, if the regularization scale was 
treated as an additional fit parameter. 
This indicates the ability of a bare resonance mass, a coupling scale, 
and regulator to compensate for underlying interactions that do not appear 
directly in the model. It also indicates that more complicated interactions 
may be effectively incorporated into the degrees of freedom of the model. 

Thus the matrix Hamiltonian model, which uses an effective field theory 
style of approach, may indeed act as a guide to lattice QCD calculations 
in identifying the salient features derived from finite-volume effects, 
and the impact of the leading-order chiral contributions near a resonance. 
Though lattice QCD calculations are typically performed at pion masses 
heavier than the physical value, the avoided level crossing feature is 
expected to occur on the lattice when the pion-nucleon threshold appears 
below the $\Delta$ resonance. 
After considering some ideal pseudodata, 
the matrix Hamiltonian approach is well equipped to describe the behaviour of 
actual lattice QCD results. 

\begin{acknowledgements}
\vspace{-2mm}
This research 
was supported by the Australian Research Council through 
 the ARC Centre of Excellence for Particle Physics at the 
Terascale, and through Grants No. DP110101265 (D.B.L. and R.D.Y.), 
No. FT120100821 (R.D.Y.) 
 and No. FL0992247 (A.W.T.). 
\end{acknowledgements}

\bibliographystyle{apsrev} 
\bibliography{refs}

\begin{thebibliography}{37}
\expandafter\ifx\csname natexlab\endcsname\relax\def\natexlab#1{#1}\fi
\expandafter\ifx\csname bibnamefont\endcsname\relax
  \def\bibnamefont#1{#1}\fi
\expandafter\ifx\csname bibfnamefont\endcsname\relax
  \def\bibfnamefont#1{#1}\fi
\expandafter\ifx\csname citenamefont\endcsname\relax
  \def\citenamefont#1{#1}\fi
\expandafter\ifx\csname url\endcsname\relax
  \def\url#1{\texttt{#1}}\fi
\expandafter\ifx\csname urlprefix\endcsname\relax\def\urlprefix{URL }\fi
\providecommand{\bibinfo}[2]{#2}
\providecommand{\eprint}[2][]{\url{#2}}

\bibitem[{\citenamefont{Edwards et~al.}(2011)\citenamefont{Edwards, Dudek,
  Richards, and Wallace}}]{Edwards:2011jj}
\bibinfo{author}{\bibfnamefont{R.~G.} \bibnamefont{Edwards}},
  \bibinfo{author}{\bibfnamefont{J.~J.} \bibnamefont{Dudek}},
  \bibinfo{author}{\bibfnamefont{D.~G.} \bibnamefont{Richards}},
  \bibnamefont{and} \bibinfo{author}{\bibfnamefont{S.~J.}
  \bibnamefont{Wallace}}, \bibinfo{journal}{Phys.Rev.}
  \textbf{\bibinfo{volume}{D84}}, \bibinfo{pages}{074508}
  (\bibinfo{year}{2011}), \eprint{1104.5152}.

\bibitem[{\citenamefont{Engel et~al.}(2010)\citenamefont{Engel, Lang, Limmer,
  Mohler, and Schafer}}]{Engel:2010my}
\bibinfo{author}{\bibfnamefont{G.~P.} \bibnamefont{Engel}},
  \bibinfo{author}{\bibfnamefont{C.}~\bibnamefont{Lang}},
  \bibinfo{author}{\bibfnamefont{M.}~\bibnamefont{Limmer}},
  \bibinfo{author}{\bibfnamefont{D.}~\bibnamefont{Mohler}}, \bibnamefont{and}
  \bibinfo{author}{\bibfnamefont{A.}~\bibnamefont{Schafer}}
  (\bibinfo{collaboration}{[BGR (Bern-Graz-Regensburg) Collaboration]}),
  \bibinfo{journal}{Phys.Rev.} \textbf{\bibinfo{volume}{D82}},
  \bibinfo{pages}{034505} (\bibinfo{year}{2010}), \eprint{1005.1748}.

\bibitem[{\citenamefont{Mahbub et~al.}(2012)\citenamefont{Mahbub, Kamleh,
  Leinweber, Moran, and Williams}}]{Mahbub:2010rm}
\bibinfo{author}{\bibfnamefont{M.~S.} \bibnamefont{Mahbub}},
  \bibinfo{author}{\bibfnamefont{W.}~\bibnamefont{Kamleh}},
  \bibinfo{author}{\bibfnamefont{D.~B.} \bibnamefont{Leinweber}},
  \bibinfo{author}{\bibfnamefont{P.~J.} \bibnamefont{Moran}}, \bibnamefont{and}
  \bibinfo{author}{\bibfnamefont{A.~G.} \bibnamefont{Williams}}
  (\bibinfo{collaboration}{CSSM Lattice Collaboration}),
  \bibinfo{journal}{Phys.Lett.} \textbf{\bibinfo{volume}{B707}},
  \bibinfo{pages}{389} (\bibinfo{year}{2012}), \eprint{1011.5724}.

\bibitem[{\citenamefont{Menadue et~al.}(2012)\citenamefont{Menadue, Kamleh,
  Leinweber, and Mahbub}}]{Menadue:2011pd}
\bibinfo{author}{\bibfnamefont{B.~J.} \bibnamefont{Menadue}},
  \bibinfo{author}{\bibfnamefont{W.}~\bibnamefont{Kamleh}},
  \bibinfo{author}{\bibfnamefont{D.~B.} \bibnamefont{Leinweber}},
  \bibnamefont{and} \bibinfo{author}{\bibfnamefont{M.~S.}
  \bibnamefont{Mahbub}}, \bibinfo{journal}{Phys.Rev.Lett.}
  \textbf{\bibinfo{volume}{108}}, \bibinfo{pages}{112001}
  (\bibinfo{year}{2012}), \eprint{1109.6716}.

\bibitem[{\citenamefont{Mohler}(2012)}]{Mohler:2012nh}
\bibinfo{author}{\bibfnamefont{D.}~\bibnamefont{Mohler}},
  \bibinfo{journal}{PoS} \textbf{\bibinfo{volume}{LATTICE2012}},
  \bibinfo{pages}{003} (\bibinfo{year}{2012}), \eprint{1211.6163}.

\bibitem[{\citenamefont{He et~al.}(2005)\citenamefont{He, Feng, and
  Liu}}]{He:2005ey}
\bibinfo{author}{\bibfnamefont{S.}~\bibnamefont{He}},
  \bibinfo{author}{\bibfnamefont{X.}~\bibnamefont{Feng}}, \bibnamefont{and}
  \bibinfo{author}{\bibfnamefont{C.}~\bibnamefont{Liu}},
  \bibinfo{journal}{JHEP} \textbf{\bibinfo{volume}{0507}}, \bibinfo{pages}{011}
  (\bibinfo{year}{2005}), \eprint{hep-lat/0504019}.

\bibitem[{\citenamefont{Bernard et~al.}(2008)\citenamefont{Bernard, Lage,
  Mei{\ss}ner, and Rusetsky}}]{Bernard:2008ax}
\bibinfo{author}{\bibfnamefont{V.}~\bibnamefont{Bernard}},
  \bibinfo{author}{\bibfnamefont{M.}~\bibnamefont{Lage}},
  \bibinfo{author}{\bibfnamefont{U.-G.} \bibnamefont{Mei{\ss}ner}},
  \bibnamefont{and} \bibinfo{author}{\bibfnamefont{A.}~\bibnamefont{Rusetsky}},
  \bibinfo{journal}{JHEP} \textbf{\bibinfo{volume}{0808}}, \bibinfo{pages}{024}
  (\bibinfo{year}{2008}), \eprint{0806.4495}.

\bibitem[{\citenamefont{Bernard et~al.}(2009)\citenamefont{Bernard, Hoja,
  Mei{\ss}ner, and Rusetsky}}]{Bernard:2009mw}
\bibinfo{author}{\bibfnamefont{V.}~\bibnamefont{Bernard}},
  \bibinfo{author}{\bibfnamefont{D.}~\bibnamefont{Hoja}},
  \bibinfo{author}{\bibfnamefont{U.-G.} \bibnamefont{Mei{\ss}ner}},
  \bibnamefont{and} \bibinfo{author}{\bibfnamefont{A.}~\bibnamefont{Rusetsky}},
  \bibinfo{journal}{JHEP} \textbf{\bibinfo{volume}{0906}}, \bibinfo{pages}{061}
  (\bibinfo{year}{2009}), \eprint{0902.2346}.

\bibitem[{\citenamefont{D{\"o}ring and Mei{\ss}ner}(2012)}]{Doring:2011nd}
\bibinfo{author}{\bibfnamefont{M.}~\bibnamefont{D{\"o}ring}} \bibnamefont{and}
  \bibinfo{author}{\bibfnamefont{U.-G.} \bibnamefont{Mei{\ss}ner}},
  \bibinfo{journal}{JHEP} \textbf{\bibinfo{volume}{1201}}, \bibinfo{pages}{009}
  (\bibinfo{year}{2012}), \eprint{1111.0616}.

\bibitem[{\citenamefont{D{\"o}ring et~al.}(2011)\citenamefont{D{\"o}ring,
  Mei{\ss}ner, Oset, and Rusetsky}}]{Doring:2011vk}
\bibinfo{author}{\bibfnamefont{M.}~\bibnamefont{D{\"o}ring}},
  \bibinfo{author}{\bibfnamefont{U.-G.} \bibnamefont{Mei{\ss}ner}},
  \bibinfo{author}{\bibfnamefont{E.}~\bibnamefont{Oset}}, \bibnamefont{and}
  \bibinfo{author}{\bibfnamefont{A.}~\bibnamefont{Rusetsky}},
  \bibinfo{journal}{Eur.Phys.J.} \textbf{\bibinfo{volume}{A47}},
  \bibinfo{pages}{139} (\bibinfo{year}{2011}), \eprint{1107.3988}.

\bibitem[{\citenamefont{Roca and Oset}(2012)}]{Roca:2012rx}
\bibinfo{author}{\bibfnamefont{L.}~\bibnamefont{Roca}} \bibnamefont{and}
  \bibinfo{author}{\bibfnamefont{E.}~\bibnamefont{Oset}},
  \bibinfo{journal}{Phys.Rev.} \textbf{\bibinfo{volume}{D85}},
  \bibinfo{pages}{054507} (\bibinfo{year}{2012}), \eprint{1201.0438}.

\bibitem[{\citenamefont{Giudice et~al.}(2012)\citenamefont{Giudice, McManus,
  and Peardon}}]{Giudice:2012tg}
\bibinfo{author}{\bibfnamefont{P.}~\bibnamefont{Giudice}},
  \bibinfo{author}{\bibfnamefont{D.}~\bibnamefont{McManus}}, \bibnamefont{and}
  \bibinfo{author}{\bibfnamefont{M.}~\bibnamefont{Peardon}},
  \bibinfo{journal}{Phys.Rev.} \textbf{\bibinfo{volume}{D86}},
  \bibinfo{pages}{074516} (\bibinfo{year}{2012}), \eprint{1204.2745}.

\bibitem[{\citenamefont{Kreuzer and Grie{\ss}hammer}(2012)}]{Kreuzer:2012sr}
\bibinfo{author}{\bibfnamefont{S.}~\bibnamefont{Kreuzer}} \bibnamefont{and}
  \bibinfo{author}{\bibfnamefont{H.~W.} \bibnamefont{Grie{\ss}hammer}},
  \bibinfo{journal}{Eur.Phys.J.} \textbf{\bibinfo{volume}{A48}},
  \bibinfo{pages}{93} (\bibinfo{year}{2012}), \eprint{1205.0277}.

\bibitem[{\citenamefont{Albaladejo et~al.}(2012)\citenamefont{Albaladejo,
  Oller, Oset, Rios, and Roca}}]{Albaladejo:2012jr}
\bibinfo{author}{\bibfnamefont{M.}~\bibnamefont{Albaladejo}},
  \bibinfo{author}{\bibfnamefont{J.}~\bibnamefont{Oller}},
  \bibinfo{author}{\bibfnamefont{E.}~\bibnamefont{Oset}},
  \bibinfo{author}{\bibfnamefont{G.}~\bibnamefont{Rios}}, \bibnamefont{and}
  \bibinfo{author}{\bibfnamefont{L.}~\bibnamefont{Roca}},
  \bibinfo{journal}{JHEP} \textbf{\bibinfo{volume}{1208}}, \bibinfo{pages}{071}
  (\bibinfo{year}{2012}), \eprint{1205.3582}.

\bibitem[{\citenamefont{Dudek et~al.}(2013)\citenamefont{Dudek, Edwards, and
  Thomas}}]{Dudek:2012xn}
\bibinfo{author}{\bibfnamefont{J.~J.} \bibnamefont{Dudek}},
  \bibinfo{author}{\bibfnamefont{R.~G.} \bibnamefont{Edwards}},
  \bibnamefont{and} \bibinfo{author}{\bibfnamefont{C.~E.}
  \bibnamefont{Thomas}}, \bibinfo{journal}{Phys. Rev. D 87,}
  \textbf{\bibinfo{volume}{034505}} (\bibinfo{year}{2013}), \eprint{1212.0830}.

\bibitem[{\citenamefont{Gockeler et~al.}(2012)\citenamefont{Gockeler, Horsley,
  Lage, Mei{\ss}ner, Rakow et~al.}}]{Gockeler:2012yj}
\bibinfo{author}{\bibfnamefont{M.}~\bibnamefont{Gockeler}},
  \bibinfo{author}{\bibfnamefont{R.}~\bibnamefont{Horsley}},
  \bibinfo{author}{\bibfnamefont{M.}~\bibnamefont{Lage}},
  \bibinfo{author}{\bibfnamefont{U.-G.} \bibnamefont{Mei{\ss}ner}},
  \bibinfo{author}{\bibfnamefont{P.}~\bibnamefont{Rakow}},
  \bibnamefont{et~al.}, \bibinfo{journal}{Phys.Rev.}
  \textbf{\bibinfo{volume}{D86}}, \bibinfo{pages}{094513}
  (\bibinfo{year}{2012}), \eprint{1206.4141}.

\bibitem[{\citenamefont{L{\"u}scher}(1991)}]{Luscher:1990ux}
\bibinfo{author}{\bibfnamefont{M.}~\bibnamefont{L{\"u}scher}},
  \bibinfo{journal}{Nucl.Phys.} \textbf{\bibinfo{volume}{B354}},
  \bibinfo{pages}{531} (\bibinfo{year}{1991}).

\bibitem[{\citenamefont{Jenkins and
  Manohar}(1991{\natexlab{a}})}]{Jenkins:1991ne}
\bibinfo{author}{\bibfnamefont{E.~E.} \bibnamefont{Jenkins}} \bibnamefont{and}
  \bibinfo{author}{\bibfnamefont{A.~V.} \bibnamefont{Manohar}},
  \bibinfo{journal}{Phys. Lett. B} \textbf{\bibinfo{volume}{259}},
  \bibinfo{pages}{353} (\bibinfo{year}{1991}{\natexlab{a}}),
  \bibinfo{note}{proceedings, Effective Field Theories of the Standard Model,
  Dobog{\'o}k\H{o}, Hungary, 1991, pp. 113-137}.

\bibitem[{\citenamefont{Jenkins and
  Manohar}(1991{\natexlab{b}})}]{Jenkins:1990jv}
\bibinfo{author}{\bibfnamefont{E.~E.} \bibnamefont{Jenkins}} \bibnamefont{and}
  \bibinfo{author}{\bibfnamefont{A.~V.} \bibnamefont{Manohar}},
  \bibinfo{journal}{Phys. Lett.} \textbf{\bibinfo{volume}{B255}},
  \bibinfo{pages}{558} (\bibinfo{year}{1991}{\natexlab{b}}).

\bibitem[{\citenamefont{Jenkins}(1992)}]{Jenkins:1991ts}
\bibinfo{author}{\bibfnamefont{E.~E.} \bibnamefont{Jenkins}},
  \bibinfo{journal}{Nucl. Phys.} \textbf{\bibinfo{volume}{B368}},
  \bibinfo{pages}{190} (\bibinfo{year}{1992}).

\bibitem[{\citenamefont{Labrenz and Sharpe}(1996)}]{Labrenz:1996jy}
\bibinfo{author}{\bibfnamefont{J.~N.} \bibnamefont{Labrenz}} \bibnamefont{and}
  \bibinfo{author}{\bibfnamefont{S.~R.} \bibnamefont{Sharpe}},
  \bibinfo{journal}{Phys. Rev.} \textbf{\bibinfo{volume}{D54}},
  \bibinfo{pages}{4595} (\bibinfo{year}{1996}), \eprint{hep-lat/9605034}.

\bibitem[{\citenamefont{Walker-Loud}(2005)}]{WalkerLoud:2004hf}
\bibinfo{author}{\bibfnamefont{A.}~\bibnamefont{Walker-Loud}},
  \bibinfo{journal}{Nucl. Phys.} \textbf{\bibinfo{volume}{A747}},
  \bibinfo{pages}{476} (\bibinfo{year}{2005}), \eprint{hep-lat/0405007}.

\bibitem[{\citenamefont{Wang et~al.}(2007)\citenamefont{Wang, Leinweber,
  Thomas, and Young}}]{Wang:2007iw}
\bibinfo{author}{\bibfnamefont{P.}~\bibnamefont{Wang}},
  \bibinfo{author}{\bibfnamefont{D.~B.} \bibnamefont{Leinweber}},
  \bibinfo{author}{\bibfnamefont{A.~W.} \bibnamefont{Thomas}},
  \bibnamefont{and} \bibinfo{author}{\bibfnamefont{R.~D.} \bibnamefont{Young}},
  \bibinfo{journal}{Phys. Rev.} \textbf{\bibinfo{volume}{D75}},
  \bibinfo{pages}{073012} (\bibinfo{year}{2007}), \eprint{hep-ph/0701082}.

\bibitem[{\citenamefont{Pascalutsa and
  Vanderhaeghen}(2006)}]{Pascalutsa:2005vq}
\bibinfo{author}{\bibfnamefont{V.}~\bibnamefont{Pascalutsa}} \bibnamefont{and}
  \bibinfo{author}{\bibfnamefont{M.}~\bibnamefont{Vanderhaeghen}},
  \bibinfo{journal}{Phys.Rev.} \textbf{\bibinfo{volume}{D73}},
  \bibinfo{pages}{034003} (\bibinfo{year}{2006}), \eprint{hep-ph/0512244}.

\bibitem[{\citenamefont{Young et~al.}(2003)\citenamefont{Young, Leinweber, and
  Thomas}}]{Young:2002ib}
\bibinfo{author}{\bibfnamefont{R.~D.} \bibnamefont{Young}},
  \bibinfo{author}{\bibfnamefont{D.~B.} \bibnamefont{Leinweber}},
  \bibnamefont{and} \bibinfo{author}{\bibfnamefont{A.~W.}
  \bibnamefont{Thomas}}, \bibinfo{journal}{Prog. Part. Nucl. Phys.}
  \textbf{\bibinfo{volume}{50}}, \bibinfo{pages}{399} (\bibinfo{year}{2003}),
  \eprint{hep-lat/0212031}.

\bibitem[{\citenamefont{Leinweber et~al.}(2004)\citenamefont{Leinweber, Thomas,
  and Young}}]{Leinweber:2003dg}
\bibinfo{author}{\bibfnamefont{D.~B.} \bibnamefont{Leinweber}},
  \bibinfo{author}{\bibfnamefont{A.~W.} \bibnamefont{Thomas}},
  \bibnamefont{and} \bibinfo{author}{\bibfnamefont{R.~D.} \bibnamefont{Young}},
  \bibinfo{journal}{Phys. Rev. Lett.} \textbf{\bibinfo{volume}{92}},
  \bibinfo{pages}{242002} (\bibinfo{year}{2004}), \eprint{hep-lat/0302020}.

\bibitem[{\citenamefont{Leinweber et~al.}(2005)\citenamefont{Leinweber, Thomas,
  and Young}}]{Leinweber:2005xz}
\bibinfo{author}{\bibfnamefont{D.~B.} \bibnamefont{Leinweber}},
  \bibinfo{author}{\bibfnamefont{A.~W.} \bibnamefont{Thomas}},
  \bibnamefont{and} \bibinfo{author}{\bibfnamefont{R.~D.} \bibnamefont{Young}},
  \bibinfo{journal}{Nucl.Phys.} \textbf{\bibinfo{volume}{A755}},
  \bibinfo{pages}{59} (\bibinfo{year}{2005}), \eprint{hep-lat/0501028}.

\bibitem[{\citenamefont{Hall et~al.}(2010)\citenamefont{Hall, Leinweber, and
  Young}}]{Hall:2010ai}
\bibinfo{author}{\bibfnamefont{J.}~\bibnamefont{Hall}},
  \bibinfo{author}{\bibfnamefont{D.}~\bibnamefont{Leinweber}},
  \bibnamefont{and} \bibinfo{author}{\bibfnamefont{R.}~\bibnamefont{Young}},
  \bibinfo{journal}{Phys.Rev.} \textbf{\bibinfo{volume}{D82}},
  \bibinfo{pages}{034010} (\bibinfo{year}{2010}), \eprint{1002.4924}.

\bibitem[{\citenamefont{Hall et~al.}(2011)\citenamefont{Hall, Lee, Leinweber,
  Liu, Mathur et~al.}}]{Hall:2011en}
\bibinfo{author}{\bibfnamefont{J.}~\bibnamefont{Hall}},
  \bibinfo{author}{\bibfnamefont{F.}~\bibnamefont{Lee}},
  \bibinfo{author}{\bibfnamefont{D.}~\bibnamefont{Leinweber}},
  \bibinfo{author}{\bibfnamefont{K.}~\bibnamefont{Liu}},
  \bibinfo{author}{\bibfnamefont{N.}~\bibnamefont{Mathur}},
  \bibnamefont{et~al.}, \bibinfo{journal}{Phys.Rev.}
  \textbf{\bibinfo{volume}{D84}}, \bibinfo{pages}{114011}
  (\bibinfo{year}{2011}), \eprint{1101.4411}.

\bibitem[{\citenamefont{Theberge et~al.}(1980)\citenamefont{Theberge, Thomas,
  and Miller}}]{Theberge:1980ye}
\bibinfo{author}{\bibfnamefont{S.}~\bibnamefont{Theberge}},
  \bibinfo{author}{\bibfnamefont{A.~W.} \bibnamefont{Thomas}},
  \bibnamefont{and} \bibinfo{author}{\bibfnamefont{G.~A.}
  \bibnamefont{Miller}}, \bibinfo{journal}{Phys.Rev.}
  \textbf{\bibinfo{volume}{D22}}, \bibinfo{pages}{2838} (\bibinfo{year}{1980}).

\bibitem[{\citenamefont{Thomas et~al.}(1981)\citenamefont{Thomas, Theberge, and
  Miller}}]{Thomas:1981vc}
\bibinfo{author}{\bibfnamefont{A.~W.} \bibnamefont{Thomas}},
  \bibinfo{author}{\bibfnamefont{S.}~\bibnamefont{Theberge}}, \bibnamefont{and}
  \bibinfo{author}{\bibfnamefont{G.~A.} \bibnamefont{Miller}},
  \bibinfo{journal}{Phys.Rev.} \textbf{\bibinfo{volume}{D24}},
  \bibinfo{pages}{216} (\bibinfo{year}{1981}).

\bibitem[{\citenamefont{Mongan}(1969)}]{Mongan:1969dc}
\bibinfo{author}{\bibfnamefont{T.}~\bibnamefont{Mongan}},
  \bibinfo{journal}{Phys.Rev.} \textbf{\bibinfo{volume}{178}},
  \bibinfo{pages}{1597} (\bibinfo{year}{1969}).

\bibitem[{\citenamefont{Liu et~al.}(2006)\citenamefont{Liu, Feng, and
  He}}]{Liu:2005kr}
\bibinfo{author}{\bibfnamefont{C.}~\bibnamefont{Liu}},
  \bibinfo{author}{\bibfnamefont{X.}~\bibnamefont{Feng}}, \bibnamefont{and}
  \bibinfo{author}{\bibfnamefont{S.}~\bibnamefont{He}},
  \bibinfo{journal}{Int.J.Mod.Phys.} \textbf{\bibinfo{volume}{A21}},
  \bibinfo{pages}{847} (\bibinfo{year}{2006}), \eprint{hep-lat/0508022}.

\bibitem[{\citenamefont{Bernard et~al.}(2012)\citenamefont{Bernard, Hoja,
  Mei{\ss}ner, and Rusetsky}}]{Bernard:2012bi}
\bibinfo{author}{\bibfnamefont{V.}~\bibnamefont{Bernard}},
  \bibinfo{author}{\bibfnamefont{D.}~\bibnamefont{Hoja}},
  \bibinfo{author}{\bibfnamefont{U.}~\bibnamefont{Mei{\ss}ner}},
  \bibnamefont{and} \bibinfo{author}{\bibfnamefont{A.}~\bibnamefont{Rusetsky}},
  \bibinfo{journal}{JHEP} \textbf{\bibinfo{volume}{1209}}, \bibinfo{pages}{023}
  (\bibinfo{year}{2012}), \eprint{1205.4642}.

\bibitem[{\citenamefont{Doring et~al.}(2012)\citenamefont{Doring, Mei{\ss}ner,
  Oset, and Rusetsky}}]{Doring:2012eu}
\bibinfo{author}{\bibfnamefont{M.}~\bibnamefont{Doring}},
  \bibinfo{author}{\bibfnamefont{U.}~\bibnamefont{Mei{\ss}ner}},
  \bibinfo{author}{\bibfnamefont{E.}~\bibnamefont{Oset}}, \bibnamefont{and}
  \bibinfo{author}{\bibfnamefont{A.}~\bibnamefont{Rusetsky}},
  \bibinfo{journal}{Eur.Phys.J.} \textbf{\bibinfo{volume}{A48}},
  \bibinfo{pages}{114} (\bibinfo{year}{2012}), \eprint{1205.4838}.

\bibitem[{\citenamefont{Hansen and Sharpe}(2012)}]{Hansen:2012tf}
\bibinfo{author}{\bibfnamefont{M.~T.} \bibnamefont{Hansen}} \bibnamefont{and}
  \bibinfo{author}{\bibfnamefont{S.~R.} \bibnamefont{Sharpe}},
  \bibinfo{journal}{Phys.Rev.} \textbf{\bibinfo{volume}{D86}},
  \bibinfo{pages}{016007} (\bibinfo{year}{2012}), \eprint{1204.0826}.

\bibitem[{\citenamefont{Blatt and Weisskopf}(1952)}]{Blatt}
\bibinfo{author}{\bibfnamefont{J.~M.} \bibnamefont{Blatt}} \bibnamefont{and}
  \bibinfo{author}{\bibfnamefont{W.~F.} \bibnamefont{Weisskopf}},
  \emph{\bibinfo{title}{Theoretical Nuclear Physics}} (\bibinfo{publisher}{John
  Wiley \& Sons, New York}, \bibinfo{year}{1952}).

\end{thebibliography}

\end{document}